\documentclass[envcountsame]{llncs}
\usepackage{amssymb}
\usepackage{amsmath}
\usepackage{verbatim}
\usepackage[usenames]{color}

\newtheorem{thm}{Theorem}

\newtheorem{la}[thm]{Lemma}
\newtheorem{cor}[thm]{Corollary}

 \newtheorem{rmktemp}[thm]{Remark}
\newenvironment{rmk}{\begin{rmktemp}\normalfont}{\end{rmktemp}}
\newtheorem{convtemp}[thm]{Convention}
\newenvironment{conv}{\begin{convtemp}\normalfont}{\end{convtemp}}

\newenvironment{ls}{\begin{itemize}}{\end{itemize}}

\newenvironment{pf}{\begin{proof}}{\end{proof}}

\newcommand{\bbb}[1]{\ensuremath{\mathbb {#1}}}

\newcommand{\emp}{\varnothing}
\renewcommand{\phi}{\varphi}

\newcommand{\sq}[1]{\ensuremath{\langle#1\rangle}}

\newcommand{\notarrow}{\kern .42em\not\kern -.42em\longrightarrow}
\newcommand{\F}{{\cal F}}
\newcommand{\LET}{\mbox{LET}}
\newcommand{\LFP}{\mbox{LFP}}
\renewcommand{\th}{\ensuremath{{}^{\text{th}}}}
\newcommand{\THEN}{\mbox{THEN}}

\newcommand{\Apply}{\text{Apply}}
\newcommand{\Arity}{\text{Arity}}
\newcommand{\Assgt}{\text{Assgt}}
\newcommand{\Change}{\text{Change}}
\newcommand{\Cat}{\text{Cat}}
\newcommand{\IndAsrt}{\text{IndAsrt}}
\newcommand{\List}{\text{List}}
\newcommand{\RenameAway}{\text{RenameAway}}
\newcommand{\Term}{\text{Term}}
\newcommand{\Sat}{\text{Sat}}
\newcommand{\Val}{\text{Val}}

\newcommand{\noprint}[1]{\relax}

\title{One Useful Logic\\ That Defines Its Own Truth}
\author{Andreas Blass\inst{1} \and Yuri Gurevich\inst{2}}
\institute{Math Dept, University of Michigan,
Ann Arbor, MI 48109, USA
\and
Microsoft Research, One Microsoft Way, Redmond, WA 98052, USA}

\begin{document}
\maketitle

\begin{abstract}
  Existential fixed point logic (EFPL) is a natural fit for some
  applications, and the purpose of this talk is to attract attention to
  EFPL.  The logic is also interesting in its own right as it has
  attractive properties.  One of those properties is rather unusual:
  truth of formulas can be defined (given appropriate syntactic
  apparatus) in the logic.  We mentioned that property elsewhere, and
  we use this opportunity to provide the proof.
\end{abstract}

\begin{quote}\raggedleft\small\it
Believe those who are seeking the truth. Doubt those who find it.\\[1ex]
---Andr\'{e} Gide
\end{quote}

\section{Introduction}

First-order logic lacks induction but first-order formulas can be used to define the steps of an induction.  Consider a first-order (also called elementary) formula $\phi(P,x_1,\ldots,x_j)$ where a $j$-ary relation $P$ has only positive occurrences. The formula may contain additional individual variables, relation symbols, and function symbols.  In every structure whose vocabulary is that of $\phi$ minus the symbol $P$ and where the additional individual variables are assigned particular values, we have an operator
\[
 \Gamma(P) = \{ \bar{x}:\ \phi(P,\bar{x}) \}.
\]
A relation $P$ is a \emph{closed point} of $\Gamma$ if $\Gamma(P) \subseteq P$, and $P$ is a \emph{fixed point} of $\Gamma$ if $\Gamma(P) = P$.  Since $P$ has only positive occurrences in $\phi(P,\bar{x})$, the operator is monotone: if $P\subseteq Q$ then $\Gamma(P)\subseteq\Gamma(Q)$.  By the Knaster-Tarski Theorem, $\Gamma$ has a least fixed point $P^*$ which is also the least closed point of $\Gamma$ \cite{tarski}.

There is a standard way to construct $P^*$ from the empty set by iterating the operator $\Gamma$.  Let $P^0 = \emptyset$, $P^{\alpha+1} = \Gamma(P^\alpha)$ and $P^\lambda = \bigcup_{\alpha<\lambda} P^\alpha$ if $\lambda$ is a limit ordinal.  There is an ordinal $\alpha$ such that $P^\alpha = P^{\alpha+1} = P^*$.  The least such ordinal $\alpha$ is the \emph{closure ordinal} of the iteration.  Such elementary inductions have been extensively studied in logic \cite{ynm,aczel}.

Notice that we have not really used the fact that $\phi(P,\bar{x})$ is first-order.  One property of $\phi(P,\bar{x})$ that we used was that $\phi(P,\bar{x})$ is monotone in $P$, that is that, in every structure of the appropriate vocabulary with fixed values for the additional individual variables, $\Gamma$ is a monotone operator. $\phi(P,\bar{x})$ could be e.g.\ a second-order formula monotone in $P$.

The least fixed point $P^*$ can be denoted $\LFP_{P,\bar{x}}\phi(P,\bar{x})$ and viewed as a $j$-ary relation, so that $[\LFP_{P,\bar{x}}\phi(P,\bar{x})](y_1,\ldots,y_j)$ functions semantically as a formula.  This observation gives rise to an idea to use LFP as a new formula constructor, in addition to propositional connectives and quantifiers.  Aho and Ullman \cite{au} indeed suggested to enrich first-order logic with the LFP constructor.  The new logic became known as FOL+LFP.

Model checking is polynomial time for any FOL+LFP formula $\psi$. In other words, it can be checked in time polynomial in the size of a finite structure $X$ of the vocabulary of $\psi$ whether $X$ with some values for the free individual variables of $\psi$ is a model of $\psi$.  Immerman \cite{immerman} and Vardi \cite{vardi} proved that, over ordered finite structures, the converse is true: every property that model checks in polynomial time is expressible in FOL+LFP.  In that sense, FOL+LFP captures polynomial time.

Existential fixed point logic (EFPL) is essentially an extension of the existential fragment of first-order logic with the LFP construct.  It does not have the universal quantifier and lacks means to simulate universal quantification; see the definition of EFPL in the next section.  As far as we know, it was first introduced --- in a different guise --- by Chandra and Harel \cite{ch} in the context of database theory where vocabularies are relational, that is, consist of relation symbols and constants and do not have function symbols of positive arity.  Chandra and Harel observed that relational EFPL is equi-expressive with Datalog, a popular database query language.

Existential fixed point logic (EFPL) was further developed by the present authors in \cite{efp}; see Section~\ref{sec:properties}.  The motivation came from program verification.  We noticed that EFPL was appropriate for formulating pre- and post-conditions in Hoare's logic of asserted programs \cite{hoare}.  In particular, the heavy expressivity hypothesis needed for Cook's completeness theorem \cite{cook} in the context of first-order logic is automatically satisfied in the context of EFPL.

More recent developments include a deductive system for EFPL introduced by Compton \cite{compton} and a normal form for EFPL formulas discovered by Grohe \cite{grohe}, who also studied connections between EFPL and other logics.  One of the present authors found connections with topos theory and showed that these connections imply some of the other, previously known, nice properties of EFPL \cite{ab-topoi1,ab-topoi2}.  The other of the present authors, together with Neeman, applied a logic equivalent to EFPL, called liberal Datalog, to develop a powerful authorization language \cite{dkal}; the equivalence between liberal Datalog and EFPL is shown in detail in \cite{g193}.

In this note, we recall the definition and known properties of EFPL, and then we prove that the truth definition of EFPL formulas can be given in EFPL.

\begin{rmk}
Nikolaj Bj{\o}rner \cite{nikolaj} observed that writing a truth definition for EFPL in EFPL is related to writing an interpreter for EFPL in EFPL. Indeed. But the interesting issue is out of scope here, in this paper, and will have to be addressed elsewhere.
\end{rmk}

\section{Existential fixed-point logic: Definition}
\label{sec:definitions}

As indicated in the introduction, existential fixed-point logic differs from first-order logic in two respects, the absence of the universal quantifier and the presence of the least-fixed-point operator. Both of these deserve some clarification.

First we define existential logic EL.  Notice that mere removal of the
universal quantifier $\forall$ has no real effect on first-order
logic, since $\forall x\,\phi$ can be expressed as $\neg\exists
x\,\neg\phi$.  To correctly define the existential fragment of
first-order logic, one must prevent such surreptitious reintroduction
of the universal quantifier.  A traditional way to do that is to
insist that all formulas have the prenex existential form $\exists x_1
\ldots \exists x_n \phi(x_1,\ldots,x_n)$ where $\phi$ is
quantifier-free.

But there is an alternative and more convenient form of the existential fragment proposed in \cite{efp}: Allow as propositional connectives only conjunction, disjunction, and negation; use only the existential quantifier; and apply negation only to atomic formulas.  It is easy to see that every formula in this alternative fragment is equivalent to one in prenex existential form, and the other way round.

With an eye on the forthcoming introduction of recursion, we stipulate that all relation symbols are divided into two categories: \emph{negatable} and \emph{positive}.  And we restrict further the use of negation in the alternative existential fragment of first-order logic: negation can be applied only to atomic formulas with negatable relation symbols.  The resulting fragment of first-order logic will be called \emph{existential logic} and denoted EL.

Now we extend existential logic by adding a new formula constructor. As usual, formulas are built by induction from atomic formulas by means of formula constructors.  In the case of EFPL, the formula constructors are those of existential logic --- the three propositional connectives and the existential quantifier --- and one additional LET-THEN constructor that is used to construct induction assertions.  We explain how the new constructor works.

Let $\F$ be the collection of formulas constructed so far.  A \emph{logic rule} has the form $P(x_1,\ldots,x_j)\leftarrow \delta(P,x_1,\ldots,x_j)$ where $P$ is a positive relation symbol of arity $j$, the $x_i$'s are distinct variables and $\delta$ is any formula in $\F$.  We wrote $\delta$ as $\delta(P,x_1,\ldots,x_j)$ to emphasize that it is allowed to contain the relation symbol $P$ and the individual variables $x_1,\ldots,x_j$, but it may also contain additional individual variables, relation symbols, and function symbols.  $P$ is the \emph{head symbol} of the rule and $\delta$ is its \emph{body}. Note that the arrow $\leftarrow$ in a logic rule is not the (reverse) implication connective but a special symbol whose only use, in our syntax, is in forming logic rules.  A \emph{logic program} is a finite collection of logic rules. (To write a program as text, one needs to order its rules, but the choice of ordering will never matter.) To be compatible with \cite{efp}, we require that different rules have different head symbols; we could remove this restriction. If $\Pi$ is a program and $\phi$ is a formula in $\F$ then
\[
 \LET\ \Pi\ \THEN\ \phi
\]
is an EFPL formula, an \emph{induction assertion}.
If $P(x_1,\ldots,x_j) \leftarrow  \delta$ is a rule in $\Pi$ then all occurrences of the variables $x_1,\ldots,x_j$ in the rule are bound occurrences in the induction assertion. And $P$ is a bound relation variable in the induction assertion.

In general, an occurrence of an individual variable $v$ in a formula $\psi$ is bound if it belongs to a subformula of the form $\exists v\, \alpha$ or to a rule of the form $P(\ldots,v,\ldots) \leftarrow \delta$; otherwise the occurrence is free. The free individual variables of $\psi$ are those with free occurrences in $\psi$. An occurrence of relation symbol $P$ in $\psi$ is bound if it belongs to subformula LET $\Pi$ THEN $\phi$ of $\psi$ and $P$ is a head symbol of $\Pi$; otherwise the occurrence is free. The vocabulary of $\psi$ consists of all the function symbols in $\psi$ and all relation symbols with free occurrences in $\psi$.

It remains to define the semantics of the induction assertion $\psi$ = LET $\Pi$ THEN $\phi$.  To simplify the exposition, we presume that the program $\Pi$ consists of two rules, $P(x_1,\ldots,x_j)\leftarrow \alpha$ and $Q(y_1,\ldots,y_k)\leftarrow \beta$.  In every structure of the vocabulary of $\psi$ with fixed values for the free individual variables of $\psi$, the program gives rise to an operator
\[
 \Gamma(P,Q) \leftarrow (\{\bar{x}:\ \alpha\},\{\bar{y}:\ \beta\}).
\]
Since $P$ and $Q$ are positive relation symbols, $\Gamma$ is monotone and thus has a least fixed point $(P^*,Q^*)$.  To evaluate $\psi$, evaluate $\phi$ using $P^*$ and $Q^*$ as the values of relations $P$ and $Q$.

\section{EFPL: Some properties}
\label{sec:properties}

We describe some properties of EFPL. The default reference is \cite{efp}.

\subsection*{Capturing polynomial time}

EFPL captures polynomial time computability over structures of the form $\{0,1,\dots,n\}$ with (at least) the successor relation and names for the endpoints.  In contrast to the corresponding result for FOL+LFP mentioned above, we use the successor relation here rather than the ordering relation $<$.  In fact, both proofs depend on the successor relation rather than the order, but in FOL one can define successor in terms of order (but not vice versa), whereas in EFPL one can define order in terms of successor (but not vice versa).

\subsection*{Validity is r.e.\ complete}

The set of logically valid EFPL formulas is recursively enumerable (in short r.e.). Furthermore, every r.e. set reduces, by means of a recursive function, to the set of valid EFPL formulas.  Thus the set of valid EFPL formulas is a complete r.e.\ set.

\subsection*{Satisfiability is r.e.\ complete}

The set of satisfiable EFPL formulas is a complete r.e. set.

\subsection*{Finite validity is co-r.e.\ complete}

The set of EFPL formulas that hold in all finite structures is a complete co-r.e.\ set.  In other words, the set of EFPL formulas $\psi$ such that $\psi$ fails in some finite structure is a complete r.e.\ set.

\subsection*{Finite model property}

When an EFPL formula $\psi$ is satisfied in a structure $X$, this fact depends on only a finite part of the structure $X$.  More precisely, there is a finite subset $D$ of the elements of $X$ such that $\psi$ is satisfied in every structure $X'$ of the vocabulary of $X$ that coincides with $X$ on $D$.  Note that $X'$ can be always chosen to be finite. If we allow basic functions of a structure to be partial, then the property in question can be formulated in a particularly simple way: If an EFPL formula is satisfied in a structure then it is satisfied in a finite substructure.

\subsection*{No transfinite induction is needed}

The closure ordinal of any monotone induction
 \[
   P \mapsto \{\bar{x}: \phi(P,\bar{x}) \},
\]
where $\phi$ is EFPL is at most $\omega$, the first infinite ordinal. The definition of the closure ordinal generalizes in a straightforward way to simultaneous monotone induction. The closure ordinal of the induction given by any logic program is at most $\omega$.

\subsection*{Truth is preserved by homomorphisms}

Truth of EFPL formulas is preserved by homomorphisms.  Here a homomorphism is a function $h$ from one structure to another such that
\begin{itemize}
\item $h$ commutes with (the interpretations of) function symbols,
\item $P(a_1,\ldots,a_j)$ implies $P(ha_1,\ldots,ha_j)$\\ for every positive relation symbol $P$ of any arity $j$, and
\item $P(a_1,\ldots,a_j)$ if and only if $P(ha_1,\ldots,ha_j)$\\ for every negatable relation symbol $P$ of any arity $j$.
\end{itemize}

\subsection*{EFPL $\cap$ FOL $\subseteq$ EL\quad}

If an EFPL formula $\phi$ is expressible in first-order logic then $\phi$ is equivalent to an existential formula. Only a limited form of this result survives in finite model theory. If an EFPL formula $\phi$ without function symbols and without negations is equivalent, on finite structures, to a first-order formula, then $\phi$ is equivalent, on finite structures, to an existential formula without negations \cite{ajtai,rossman}. This fails even if $\phi$ has no function symbols and only the equality relation is negatable \cite[Section~10]{ajtai}.

\section{Prerequisites for truth}

Our objective in the rest of the article is to show that EFPL can formalize its own truth definition.  That is, we shall define, in EFPL with suitable vocabulary, truth of EFPL sentences (that is formulas with no free variables) of the same vocabulary.
We use the term predicate to mean a relation symbol or a relation depending on the context.

Since sentences are built from subformulas that may have free variables, we shall actually define the slightly more general concept of satisfaction of formulas by assignments of values to the free variables.  The need to define the more general notion of satisfaction of formulas in order to obtain truth for sentences is familiar from first-order logic.%
\footnote{A few authors, notably Shoenfield \cite{shoenfield}, define truth directly.  To do so, they expand the vocabulary by adding constants for all elements of the structure under consideration, and instead of assigning values to variables they substitute constants for variables.  We could have used this approach for EFPL, but we chose to parallel the more widely used approach in FOL, via satisfaction.}
A new complication, of the same general nature, arises in EFPL.  The bound predicates of a sentence $\phi$ are free in some subformulas of $\phi$.  We should define satisfaction of $\phi$ in a structure whose vocabulary does not include those predicates.  But the definition will pass through subformulas of $\phi$ whose satisfaction will depend on the interpretations of those predicates. As a result, we need to define satisfaction of $\phi$ in a context that includes not only a structure (for the vocabulary of $\phi$) and an assignment of values to the free variables of $\phi$ (as in FOL) but also the logic rules that provide the meaning of all other predicates that occur in $\phi$ --- or that occur in the bodies of those rules.

Let $\Upsilon$ be a vocabulary and $X$ a structure of vocabulary $\Upsilon$. Any predicate that does not occur in $\Upsilon$ will be called an \emph{extra predicate}. We shall define satisfaction in $X$ for $\Upsilon$-formulas.   Requirements will be imposed shortly on $\Upsilon$ and $X$, but for now $\Upsilon$ is just some vocabulary and $X$ some $\Upsilon$-structure. We intend to define, in EFPL, a ternary predicate Sat such that, when
\begin{ls}
  \item the value of its first argument is a formula $\phi$, of vocabulary $\Upsilon$ plus (possibly) some extra predicates,
  \item the value of its second argument is a logic program $\Pi$ whose head predicates include all extra predicates that occur in $\phi$ or $\Pi$, and
  \item the value of its third argument is an assignment $s$ of elements of $X$ to (at least) all individual variables that are free in $\phi$ or in $\Pi$,
\end{ls}
then the truth value of $\Sat(\phi,\Pi,s)$ in $X$ is the same as the truth value, in $X$, of $\phi$ with values for its variables given by $s$ and with the extra predicates interpreted by the least fixed point of (the monotone operator defined by) $\Pi$.

Furthermore, we do not intend to use any clever tricks in our definition of Sat.  It will be a formalization of the explanation given above (and in \cite{efp}) of the meaning of EFPL formulas.  The point of this work is to show that this formalization can be carried out in EFPL itself.

For all this to make sense, the structure $X$ must contain the formulas $\phi$ of EFPL, the logic programs $\Pi$, and the assignments $s$.  Furthermore, the vocabulary must be adequate to express the basic syntactic properties of formulas and to allow basic constructions of assignments, rules, and programs.  We do not, however, wish to specify the exact syntactic nature of formulas --- for example, are they sequences of symbols, or are they parse trees, or are they G\"odel numbers?  Our work is independent of such details.  So we shall merely assume that certain notions (e.g., the operation of forming the conjunction of two formulas) are expressible; the details of how they are expressed (and which notions are primitive and which are derived) are irrelevant.%
\footnote{We shall occasionally indicate how certain notions can be defined from others in EFPL. Those indications can help to reduce the assumptions needed about $\Upsilon$.}

In the rest of this section, we list what we require of our vocabulary $\Upsilon$ and structure $X$, occasionally adding some comments about the reasons for particular requirements.

$\Upsilon$ should be finite.  The reason is that the definition of satisfaction must, in the clauses for atomic formulas, use all the relation and function symbols of $\Upsilon$.

The equality predicate should be negatable.  The reason is that the notion of EFPL formula requires some things to be distinct, for example the variables in the head of a rule and the head symbols of different rules in a program.

$X$ should contain a copy \bbb N of the natural numbers, and $\Upsilon$ should have a constant symbol for 0 and a unary function symbol $S$ for successor.  \bbb N itself, as a unary relation, is definable:
\[
\bbb N(x) :\equiv \text{LET }N(z)\leftarrow z=0\ \lor\ \exists y\,(N(y)\land z=S(y))\text{ THEN }N(x).
\]
We could also define addition and multiplication as ternary relations, and the ordering, and similarly for other primitive recursive functions and relations.

We need \bbb N primarily to index elements of lists, for example the list of terms that serves as the arguments of a relation or function symbol.  Since $\Upsilon$ is finite, we could handle the argument lists of its own relation and function symbols in an ad hoc manner, without a general notion of natural number or of list.  But EFPL imposes no bound on the arities of the head symbols of logic rules, so atomic formulas can involve arbitrarily long argument lists, and natural numbers are needed for treating these.

Although EFPL does not allow universal quantification in general, it
can simulate universal quantification over finite initial segments of
\bbb N, as shown by the following lemma from \cite{efp}.

\begin{la}   \label{bdd-all-N}
  For any EFPL formula $\phi(x)$, there is an EFPL formula $\psi(y)$ equivalent, for all $y\in\bbb N$, to $(\forall x<y)\,\phi(x)$.
\end{la}

\begin{pf}
The most natural choice of $\psi(y)$ describes a search from 0 up to $y$:\\[1ex]
\begin{minipage}{\textwidth}
\mbox{}\quad
LET $K(x)\leftarrow x=0\ \lor\ \exists w\, \big( x=S(w)\ \land\ K(w)\ \land\ \phi(w) \big)$ THEN $K(y)$.\hfill\qed
\end{minipage}
\end{pf}

\begin{conv}
Consider the definition of \bbb N exhibited above, and notice that its
essential content is contained in the rule
\[
N(z)\leftarrow z=0\, \lor\, \exists y\,(N(y)\land z=S(y)),
\]
which makes the bound predicate symbol $N$ denote the set of natural
numbers.  The rest of the definition,
\[
\bbb N(x) :\equiv \text{LET }\dots\text{ THEN }N(x),
\]
merely transfers this denotation to the defined notation \bbb N.
Instead of introducing a bound predicate variable $N$ to, in effect,
duplicate the desired predicate \bbb N, we could convey the same
information by writing
\[
\bbb N(z):\leftarrow z=0\, \lor\, \exists y\,(\bbb N(y)\land z=S(y)).
\]
Although this is not an EFPL formula, we adopt the convention that it
is to serve as an abbreviation of the definition of \bbb N displayed
earlier.  In general, when we write a rule with a colon before the
$\leftarrow$, it is to be interpreted as defining a formula.  Thus,
\[
\bbb P(\bar x):\leftarrow\delta(\bbb P,\bar x)
\]
means that $\bbb P(\bar x)$ is defined as the formula
\[
\text{LET }Q(\bar z)\leftarrow\delta(Q,\bar z)\text{ THEN }Q(\bar x).
\]
\end{conv}

\begin{conv}
Later, we shall also need to deal with definitions of this sort in which the body $\delta$ is a disjunction of many subformulas.  For example, our ultimate goal, the definition of Sat, will have several disjuncts, covering the different syntactic constructs of EFPL.  In such cases, it is convenient to present one disjunct (or a small number of them) at a time.  Thus, for a small example, the definition of \bbb N above could be broken into two parts:
\begin{align*}
    \bbb N(z)&;\leftarrow z=0\\
    \bbb N(z)&;\leftarrow \exists y\,(\bbb N(y)\land z=S(y)).
\end{align*}
We use a semicolon before $\leftarrow$ (instead of a colon) to indicate that the full definition involves more disjuncts. (This use of a semicolon as a partial colon is suggested by the word ``semicolon.") In general, if we write several semicolon definitions $\bbb P(\bar x);\leftarrow\delta_i$ for the same $P(\bar x)$, then they are to be understood as meaning $\bbb P(\bar x): \leftarrow \bigvee_i \delta_i$.
\end{conv}

Returning to the requirements on $X$ and $\Upsilon$, we require $X$ to contain the variables and the assignments.  The latter are finite partial functions from the variables into (the universe of) $X$. $\Upsilon$ should define a predicate Vbl for the set of variables, a constant symbol $\emp$ for the empty assignment, and a ternary function symbol Modify for the function defined as follows: Given an assignment $s$, a variable $v$, and an element $a$ of $X$, $\text{Modify}(s,v,a)$ is the assignment $t$ that sends $v$ to $a$ and otherwise agrees with $s$ (whether or not $a$ is in the domain of $s$).

\begin{conv}
  Here and in what follows, we use the terminology ``$\Upsilon$ should
  define a predicate for'' some relation on $X$ to mean that there
  should be an EFPL formula in vocabulary $\Upsilon$ whose truth set
  in $X$ is the desired relation.  Of course, the easiest way to
  arrange this would be for the given relation to be one of the basic
  relations of $X$, so that the required EFPL formula would be
  atomic.  But it will never matter whether the formula is atomic or
  not.

Similarly, when we ask that $\Upsilon$ should have certain function
symbols, we could weaken that to require only some terms, possibly
involving nesting of function symbols, and our proofs would be
unchanged.
\end{conv}

We also need to express ``$s$ is an assignment,'' ``$v$ is in the
domain of $s$,'' and ``$s(v)=a$,'' but we need not assume these
separately, as they are definable from $\emp$ and Modify.  They are
given, using our conventions above and the familiar convention of
(existentially) quantifying several variables at once, by
\begin{align*}
\Assgt(s)&;\leftarrow s=\emp\\
\Assgt(s)&;\leftarrow\exists t,v,a\,(\Assgt(t)\land
\text{Vbl}(v)\land s=\text{Modify}(t,v,a))\\
v\text{ inDom }s&:\leftarrow\exists t,a\,(s=\text{Modify}(t,v,a)).\\
s(v)=a&:\leftarrow\exists t\,(s=\text{Modify}(t,v,a))
\end{align*}
Note that here $s(v)=a$ is defined as a ternary relation, not as an
instance of equality.

We shall also need to have, among the elements of $X$, the relation and function symbols of $\Upsilon$ as well as the extra predicates available as head symbols of rules.  Each relation symbol $P$ or function symbol $f$ of $\Upsilon$, should be denoted by a closed term $\dot P$ or $\dot f$ of $\Upsilon$.  (We remain flexible as to what the symbols of $\Upsilon$ should be.  For example, they could be G\"odel numbers, and then their names $\dot P$ and $\dot f$ could be terms of the form $SS\dots S(0)$.  But there are many other options, and all will work.  Note, however, that we cannot take all the $\dot f$'s to be simple constant symbols, as they would then be among the $f$'s, and there would not be enough room in a finite $\Upsilon$ for all of these names to have names.)

The extra predicates available as head symbols of rules should have specified numbers of arguments.  That is, there should be an $\Upsilon$-definable predicate Arity such that $\Arity(a,n)$ holds in $X$ (for elements $a,n\in X$) if and only if $a$ is one of these head predicate symbols and $n\in\bbb N$ is the number of its argument places.

As mentioned earlier, we shall need lists, so we require that $X$
contain all lists (i.e., finite sequences) of elements of $X$.  The
vocabulary $\Upsilon$ should contain at least the constant Nil,
denoting the empty list, and the binary function symbol Append, for
the function that lengthens a list by adding one element at the end.
Thus, for example,
\[
\sq{a,b,c} =\text{Append}(\text{Append}(\text{Append}(\text{Nil},a)
,b),c).
 \]
Other predicates and functions that we shall need for dealing with
lists can be defined in terms of Nil and Append.
\begin{align*}
  \List(l);\leftarrow&\  l=\text{Nil}\\
  \List(l);\leftarrow&\  \exists x,a\,(\List(x)\land l=\text{Append}(x,a))\\
l \text{ hasLength } n;\leftarrow&\  l=\text{Nil}\land n=0\\
l \text{ hasLength } n;\leftarrow&\ \exists x,a,m\,
  \big(l=\text{Append}(x,a)\land x\text{ hasLength }m\land n=S(m)\big)\\
(l)_i=a;\leftarrow&\  \exists x\,
  \big( x\text{ hasLength }i\land l=\text{Append}(x,a) \big)\\
(l)_i=a;\leftarrow&\  \exists x,b\,
  \big( (x)_i=a\land l=\text{Append}(x,b) \big)\\
\Cat(a,b,l);\leftarrow&\  b=\text{Nil} \land l=a\\
\Cat(a,b,l);\leftarrow&\  \exists c,x,m\, \big(\Cat(a,c,m)\ \land\\
    &\ b=\text{Append}(c,x) \land (l=\text{Append}(m,x) \big).
\end{align*}
Here $(l)_i=a$, though it looks like an equation, is really a defined ternary relation, whose meaning is that $a$ is the $i\th$ component of the list $l$, where we start counting with 0, and where the length of $l$ must be at least $i+1$ so that there is an $i\th$ term. And ``Cat" alludes to ``concatenation". If $a,b,l$ are lists and $\Cat(a,b,l)$ holds, then $l$ is the concatenation $a*b$ of $a$ and $b$.

We note the following consequence of Lemma~\ref{bdd-all-N}, allowing
universal quantification over the elements of a list.

\begin{cor}    \label{bdd-all-list}
  For any EFPL formula $\phi(x)$, there is an EFPL formula $\psi(y)$
  that holds, when the value of $y$ is a list, if and only if $\phi$
  holds of all elements of that list.  That is, $\psi(y)$ is the
  result of universally quantifying $\phi(x)$ over all elements $x$ of
  the list $y$.
\end{cor}

\begin{pf}
  Use Lemma~\ref{bdd-all-N} to express\\[1ex]
\begin{minipage}{\textwidth}
\mbox{}\qquad
$\exists n\, \big( y\text{ hasLength }n\, \land\, (\forall i<n)\,\exists z\, ((y)_i=z\, \land\, \phi(z)) \big)$. \hfill\qed
\end{minipage}
\end{pf}

It will be convenient to write $(\forall x\in y)\,\phi(x)$ for the
formula $\psi$ given by this corollary.

Finally, $X$ must contain the syntactic entities relevant to EFPL,
such as terms, logic rules, logic programs, and formulas.  The precise
nature of these entities depends on arbitrary choices of how to
represent syntax.  We require merely that some representation be
present and that $\Upsilon$ be able to describe fundamental syntactic
relationships.

First, $\Upsilon$ should have a binary function symbol Apply, used to
form a compound term $f(t_1,\dots,t_n)$ from an $n$-ary function
symbol $f$ and a list \sq{t_1,\dots,t_n} of $n$ terms, and also used
similarly to form atomic formulas $P(t_1,\dots,t_n)$.  Depending on
how syntax is represented, Apply could, for example, be simply a
pairing function, or it could be the operation of prepending an
element to a list, or it could produce a tree from a root and its
immediate subtrees, or it could be an arithmetical operation on
G\"odel numbers.

There should also be a unary function symbol Neg and binary function symbols Conj, Disj, Quant, and IndAsrt for the operations of negating a formula, forming conjunctions, forming disjunctions, forming existential quantifications, and forming induction assertions LET $\Pi$ THEN $\phi$.  The arguments of these operations are intended to be formulas, except that the first argument of Quant is the variable being quantified and the first argument of IndAsrt is the program that goes between LET and THEN.

There should also be a binary function symbol Rule for the operation building a logic rule from its head and its body.  We shall take logic programs to be (certain) lists of rules, so we do not need additional capabilities in $\Upsilon$ to handle these.  (We could have used sets of rules instead, but then $\Upsilon$ would need additional capabilities.) Finally, there is a ternary relation RenameAway such that, if $\Pi$ is a program and $\phi$ is a formula and $\RenameAway(\phi,\Pi,\phi')$ holds, then $\phi'$ is a formula obtained from $\phi$ by renaming the bound predicates of $\phi$ away from the head predicates of $\Pi$, so that the formula $\phi'$ is equivalent to $\phi$, and no head predicate of $\Pi$ is bound in $\phi'$.

This completes our requirements on $\Upsilon$ and $X$.  They can be summarized thus: EFPL syntax and basic combinatorial ingredients for EFPL semantics (like assignments) are available in $X$ and expressible in EFPL in vocabulary $\Upsilon$.

\section{Semantics of terms}

Terms are built, as in FOL, by starting with variables and iteratively
applying function symbols.  The definition is formalized as follows.
\begin{align*}
  \Term(t)&;\leftarrow \text{Vbl}(t)\\
  \Term(t)&;\leftarrow \exists l \big( t=\Apply(\dot f,l)\land
  \List(l)\land l\text{ hasLength }\hat n \land(\forall x\in
  l)\Term(x) \big).
\end{align*}
Here the second line is to be repeated for each function symbol $f$ of
$\Upsilon$, $n$ is the arity of $f$, and $\hat n$ is the numeral for
$n$, namely $SS\dots S(0)$ with $n$ occurrences of $S$.  Recall that
the universal quantification $\forall x \in l$ was introduced after
Corollary~\ref{bdd-all-list} as an abbreviation of an EFPL formula.
Recall also that $\Upsilon$ is finite, so there is no difficulty
writing the appropriate line for each $f$.

Semantically, a term gets a value (in the given structure $X$) once an assignment provides values for all the variables in $t$.  So the values of terms are given by a binary function, whose arguments are a term and an assignment.  To define it recursively, we regard this binary function as a ternary relation, and we define it as follows.
\begin{align*}
\Val(t,s,a)&;\leftarrow \text{Vbl}(t)\land\Assgt(s)\land s(t)=a\\
\Val(t,s,a)&;\leftarrow \exists l,u_0,\dots,u_{n-1},b_0,\dots, b_{n-1}\\
& \big(t=\Apply(\dot f,l)\, \land\, \List(l)\, \land\,
    l\text{ hasLength }\hat n\, \land\, \Assgt(s)\\
& \land \bigwedge_{i<n}((l)_i=u_i\, \land\, \Val(u_i,s,b_i))\,
    \land\, a=f(b_1,\dots,b_n) \big).
\end{align*}
The explanatory comments after the definition of Term apply here as
well.

\begin{rmk}
  In principle, we could do without the definition of Term.  The
  definition of Val assigns values only to terms in any case.  But it
  would do no harm if Val were defined in some extraneous cases, as
  long as it worked correctly for terms.
\end{rmk}

\section{Semantics of formulas}

As indicated earlier, the semantics of a formula involves not only the structure $X$ and an assignment $s$ but also a collection $\Pi$ of logic rules to determine the meaning of any extra predicates used in the formula but not bound by LET-THEN constructions in the formula.  Ultimately, when we deal with $\Upsilon$-formulas, there will be no such extra predicates, so $\Pi$ will be irrelevant, but in the recursive construction of an $\Upsilon$-formula (and in the recursive definition of its satisfaction), subformulas can occur that do use extra predicates.  So we shall define Sat as a ternary predicate, where the intended meaning of $\Sat(\phi,\Pi,s)$ is that the formula $\phi$ is true, in our given structure $X$, when the extra predicates are interpreted by the least fixed point of $\Pi$ and the free variables are assigned values by $s$.

The definition of Sat will have numerous clauses, according to the
last constructor used in building $\phi$, so we shall make much use of
the ``;$\leftarrow$'' convention.  This way, we can present the
clauses one (or a few) at a time and insert comments and even other
definitions between them.

We begin with the case of atomic formulas whose predicates are from $\Upsilon$.  The definition is quite analogous to the earlier definition of the values of terms.
\begin{equation}
\begin{split}
\Sat(\phi,\Pi,s)&;\leftarrow \exists l,u_0,\dots,u_{n-1},b_0,\dots,b_{n-1}\\
& \big( \phi=\Apply(\dot P,l)\, \land\, \List(l)\, \land\,
    l\text{ hasLength }\hat n\, \land\, \Assgt(s)\\
& \land
    \bigwedge_{i<n}((l)_i=u_i\, \land\, \Val(u_i,s,b_i))\,
    \land\, P(b_1,\dots,b_n) \big).
\end{split}
\end{equation}
This is to be repeated for all of the (finitely many) predicates $P$ of $\Upsilon$ with $n$ being the arity of $P$.  As before, $\hat n$ is the numeral for $n$.

The case of negated atomic formulas is almost the same; of course it
is to be repeated only for negatable $P$.
\begin{equation}
\begin{split}
\Sat(\phi,\Pi,s)&;\leftarrow \exists l,u_0,\dots,u_{n-1},b_0,\dots,b_{n-1}\\
& \big(\phi=\text{Neg}(\Apply(\dot P,l))\, \land\, \List(l)\, \land\,
    l\text{ hasLength }\hat n\, \land\, \Assgt(s)\\
  &\land
  \bigwedge_{i<n}((l)_i=u_i\, \land\, \Val(u_i,s,b_i))\, \land\,
  \neg P(b_1,\dots,b_n) \big).
\end{split}
\end{equation}

Rather than continuing with the remaining atomic formulas, those that
use extra predicates, let us first dispose of the remaining ``easy''
clauses, those not involving $\Pi$.

\begin{equation}
\begin{split}
\Sat(\phi,\Pi,s)&;\leftarrow
    \exists \alpha,\beta\, \big(\phi=\text{Conj}(\alpha,\beta)\, \land\,
    \Sat(\alpha,\Pi,s)\, \land\, \Sat(\beta,\Pi,s) \big)\\
\Sat(\phi,\Pi,s)&;\leftarrow
    \exists \alpha,\beta\, \big(\phi=\text{Disj}(\alpha,\beta)\, \land\,
    (\Sat(\alpha,\Pi,s)\, \lor\, \Sat(\beta,\Pi,s) \big)\\
\Sat(\phi,\Pi,s)&;\leftarrow
    \exists\alpha,v,a\, \big(\phi=\text{Quant}(v,\alpha)\, \land\,
    \Sat(\alpha,\Pi,\text{Modify}(s,v,a)) \big)
\end{split}
\end{equation}

This completes the easier part of the definition of Sat, the part
concerning just EL.  To complete the definition for EFPL, we must deal
carefully with programs in both of their roles --- as the second
argument of Sat and as a constituent of induction assertions.

This will require some preliminaries.  First, we need the notion of a
list with no repetitions.
\begin{align*}
\text{1-1-List}(l)&:\equiv \exists n\, \big(l\text{ hasLength }n\, \land \\
& (\forall i,j<n)\,\exists x,y\,
    ((l)_i=x\, \land\, (l)_j=y\land(i=j\, \lor\, \neg(x=y))) \big).
\end{align*}
We also need a construction that amounts to applying a unary function to each element of a list, producing a new list.  The situation is complicated by the fact that our unary functions are often given as binary relations.  We therefore adopt the following notation.  If we have defined a binary relation $R$, then we write $R^+$ for the relation defined as follows.
\begin{align*}
R^+(l,m)&:\equiv\exists n\, \big(l\text{ hasLength }n\,
    \land\,  m\text{ hasLength }n\, \land \\
& (\forall i<n)\, \exists u,v\,
    ((l)_i=u\, \land\, (m)_i=v\, \land\, R(u,v)) \big).
\end{align*}
For example, let us define HS (abbreviating ``head symbol'') by
\[
HS(r,p):\equiv\exists y,z\,(r = Rule(Apply(p,y),z)).
\]
Then when $\Pi$ is a list of rules, $HS^+(\Pi,m)$ means that $m$ is the
list of their head symbols.  One of the requirements for a program is
that this list $m$ be one-to-one, so there will be a clause $\exists
m\, (HS^+(\Pi,m)\land\text{1-1-List}(m))$ in the definition of
program.

We shall also use the plus-notation with a parameter.  Specifically,
we think of $\Val(u,s,b)$ as the graph of a function $u\mapsto
b$ with $s$ fixed, so the plus-notation makes $\Val^+(\bar
u,s,\bar b)$ the relation between a list of terms and their values, all
for the same assignment $s$.  We refrain from writing out the
definition, since it's just like the definition of $R^+$ above, with
the extra argument $s$ inserted into both $R$ and $R^+$.

We need an improved version of the function Modify, to modify an
assignment by mapping all the variables in a list $l$ to the
corresponding values in another list $q$ (of the same length).
\begin{align*}
\Change(s,l,q,r)&;\leftarrow
    l=\text{Nil}\, \land\, q=\text{Nil}\, \land\, s=r\\
\Change(s,l,q,r)&;\leftarrow\exists l',q',r',v,a\,
    \big (l=\text{Append}(l',v)\, \land\, q=\text{Append}(q',a)\\
& \land\, \Change(s,l',q',r')\, \land\, r=\text{Modify}(r',v,a) \big).
\end{align*}

With these preliminaries, we can write down the definition of
satisfaction for atomic formulas that begin with one of the extra
predicates.  The idea is to find, in $\Pi$, the rule having
that symbol as its head symbol, and to use the body of that rule as
the criterion of truth for our atomic formula.  It will be useful
later to make sure that the $\Pi$ in the second argument place of Sat
has no repeated head symbols, so we include that in the definition.
\begin{equation}
\begin{split}
  \Sat(\phi,\Pi,s)&;\leftarrow\exists p,t,k,i,m,l,r,q,\delta\\
  & \big(\phi=\Apply(p,t)\, \land\, t\text{ hasLength }k\, \land\,
    \Arity(p,k)\ \land\\
  & (\forall x\in t)\ \text{Term}(x)\, \land\, \text{HS}^+(\Pi,m)\, \land\,
    \text{1-1-List}(m)\ \land\\
  & (\Pi)_i=\text{Rule}(\Apply(p,l),\delta)\, \land\, \text{1-1-List}(l)\ \land\\
  & l\text{ hasLength }k\, \land\, (\forall x\in l)\,\text{Vbl}(x)\, \land\,
    \Val^+(t,s,q)\ \land\\
  &  \Change(s,l,q,r)\, \land\, \Sat(\delta,\Pi,r) \big).
\end{split}
\end{equation}
In prose, the essential part of this says that $\phi$ has the form $p(\bar t)$ for an extra predicate of arity $k$, with $\bar t$ being a $k$-tuple of terms; that $\Pi$ contains a rule $p(\bar l)\leftarrow\delta$ with head $p$, $\bar l$ being a $k$-tuple of distinct variables; and that $\delta$ is satisfied by the assignment $r$ obtained from $s$ by replacing each of the variables in the list $\bar l$ by the value of the corresponding element of $\bar t$.  This replacement amounts, intuitively, to taking the definition of $p(\bar l)$ as $\delta(\bar l)$ and applying it to $p(\bar t)$, the terms $\bar t$ replacing the variables $\bar l$.  Instead of doing a syntactic substitution of $\bar t$ for $\bar l$ in $\delta$, we have made the corresponding semantic change, assigning to the variables in $\bar l$ the values of the terms in $\bar t$.

It may seem strange that this clause in the definition of Sat says
nothing about iterating the operator defined by $\delta$.  After all,
$p$ should be interpreted as the least fixed point of that operator.
But the desired iteration is automatically accomplished by the
iteration involved in the definition of Sat.  That is, if $p$ occurs
in $\delta$, then the true instances of $p$ can contribute to the
true instances of $\delta$ and can thereby contribute to additional
true instances of $p$.

We must still provide the clause for induction assertions in our definition of Sat.  Fortunately, this is relatively easy, since iteration is already implicitly done in the preceding clause.
\begin{equation}
\begin{split}
\Sat(\phi,\Pi,s)\ &;\leftarrow\ \exists \phi',\Sigma,\alpha,\Theta\\
& \big( \RenameAway(\phi,\Pi,\phi')\, \land\, \phi'=\IndAsrt(\Sigma,\alpha)\\
& \land\, \Cat(\Pi,\Sigma,\Theta)\, \land\, \Sat(\alpha,\Theta,s) \big).
\end{split}
\end{equation}

Here $\phi'$ is equivalent to $\phi$ and so $\Sat(\phi,\Pi,s)$ should be equivalent to $\Sat(\phi',\Pi,s)$. Further, $\phi'$ = LET $\Sigma$ THEN $\alpha$, and no head predicate of $\Pi$ is bound in $\phi'$. It follows that the head predicates of $\Pi$ are disjoint from the head predicates of $\Sigma$, so that the concatenation $\Theta$ of $\Pi$ and $\Sigma$ is a legitimate program. Accordingly $\Sat(\phi',\Pi,s)$ should be equivalent to $\Sat(\alpha,\Theta,s)$.

That concludes the definition of $\Sat(\phi,\Pi,s)$. It is easy to see that it works as intended. In the case when $\phi$ is a sentence and when both $\Pi$ and $s$ are empty, $\Sat(\phi,\Pi,s)$ holds in the structure $X$ if and only $\phi$ does.

\end{document}